\documentclass[aps, twocolumn, prb, citeautoscript, superscriptaddress, amsmath, 8pt]{revtex4-2}

\usepackage{graphicx}
\usepackage{gensymb}
\usepackage{color}
\usepackage[dvipsnames]{xcolor}
\usepackage{float}
\usepackage{upgreek}
\usepackage{bm}
\usepackage{amsmath,amssymb}
\usepackage[colorlinks=true,citecolor=blue,linkcolor=blue]{hyperref}

\begin{document}
\title{Omnidirectional Magnetic Imaging of Magnetic Anisotropy and Phase Transitions}

\author{Alexander J. Healey}
\email{alexander.healey2@rmit.edu.au}
\affiliation{Department of Physics, School of Science, RMIT University, Melbourne, VIC, Australia}

\author{Kaijian Xing}
\affiliation{School of Science, RMIT University, Melbourne, VIC, Australia}

\author{Weiyao Zhao}
\affiliation{Department of Materials Science \& Engineering, Monash University, Clayton VIC 3800, Australia}

\author{Islay O. Robertson}
\affiliation{Department of Physics, School of Science, RMIT University, Melbourne, VIC, Australia}

\author{Hark Hoe Tan}
\affiliation{ARC Centre of Excellence for Transformative Meta-Optical Systems, Department of Electronic Materials Engineering, Research School of Physics, The Australian National University, Canberra, ACT 2600, Australia}

\author{Mehran Kianinia}
\affiliation{School of Mathematical and Physical Sciences, University of Technology Sydney, Ultimo, New South Wales 2007, Australia}
\affiliation{ARC Centre of Excellence for Transformative Meta-Optical Systems, Faculty of Science, University of Technology Sydney, Ultimo, New South Wales 2007, Australia}

\author{Igor Aharonovich}
\affiliation{School of Mathematical and Physical Sciences, University of Technology Sydney, Ultimo, New South Wales 2007, Australia}
\affiliation{ARC Centre of Excellence for Transformative Meta-Optical Systems, Faculty of Science, University of Technology Sydney, Ultimo, New South Wales 2007, Australia}

\author{Jean-Philippe Tetienne}
\affiliation{Department of Physics, School of Science, RMIT University, Melbourne, VIC, Australia}

\author{Julie Karel}
\affiliation{Department of Materials Science \& Engineering, Monash University, Clayton VIC 3800, Australia}

\author{David A. Broadway}
\email{david.broadway@rmit.edu.au}
\affiliation{Department of Physics, School of Science, RMIT University, Melbourne, VIC, Australia}

\begin{abstract}
Micron scale imaging of magnetic fields is an important tool for understanding the evolution of magnetism through phase transitions and as a result of interactions inside of heterostructures.
However, most imaging platforms, like the nitrogen-vacancy (NV) centre in diamond, are restricted to applying magnetic fields along the quantisation axis of the quantum sensor. 
This greatly restricts the utility of these systems for exploring materials that emit strong fields or exhibit variable response with respect to the applied field direction.  
Here we explore an alternative approach using weakly coupled spin-pairs in hBN that exhibit a spin-1/2-like behaviour and an isotropic response to magnetic field. 
We demonstrate that the spin-pair system can operate in the presence of strong fields from a thin film magnet which were incompatible with NV diamond imaging even with applied fields along the quantisation axis.
Further, we demonstrate that using this platform allows for imaging with an arbitrary applied magnetic field direction, allowing us to probe the anisotropy and spin-reorientation transition in the ferrimagnet TbMn$_6$Sn$_6$. 
Finally, we propose an improved geometry for imaging small anisotropy contributions such as crystalline anisotropy. 
These results demonstrate how this or similar spin-1/2 systems might be used for imaging magnetic materials that are incompatible with other techniques despite the reduction in sensitivity compared with NV in diamond imaging.  
\end{abstract}

\maketitle 
\section{Introduction}
\begin{figure}
    \centering
    \includegraphics[width=\linewidth]{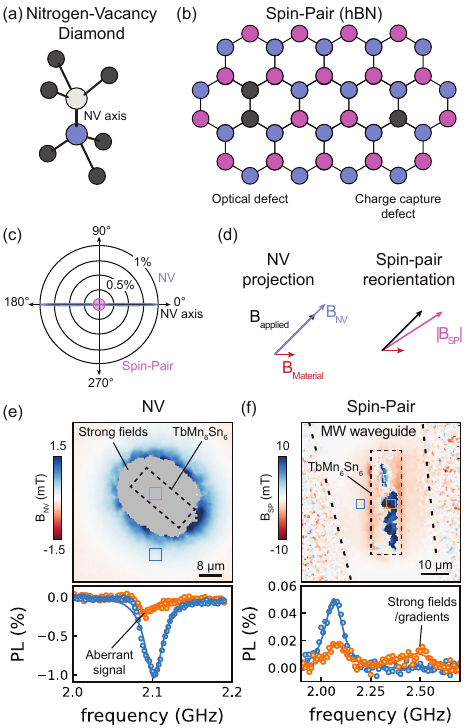}
    \caption{Introduction to spin-pair imaging.
    (a, b) Illustration of the crystal structure of (a) the NV centre in diamond and (b) the optical spin-pair in hBN. 
    (c) Comparison of the spin contrast of the NV and spin-pair for different magnetic field directions with a magnitude of $|B| = 75$\,mT. 
    (d) Illustration of the vector magnetic field components projection onto the NV quantisation axis and the reorientation of the spin-pair to match the combined field. 
    (e) Magnetic image taken with an ensemble of NV centres in the presence of a strong magnetic material that quenches the spin contrast near it (top panel) with ODMR spectra from the marked positions (bottom panel). 
    (f) Similar magnetic sample to the one imaged in (e) but now with the spin-pair in hBN which exhibits a much larger dynamic range to strong external fields. Taken at $T = 250$~K with an OOP field of $B=75$~mT.
    }
    \label{fig:intro}
\end{figure}

Imaging techniques based on quantum sensors embedded within solid state materials are powerful tools for understanding condensed matter systems~\cite{casola2018, christensen2024}. The gold standard system is the nitrogen-vacancy (NV) centre in diamond (Fig.~\ref{fig:intro}a) due to its excellent room temperature properties that enable a range of possible measurement schemes~\cite{rovny2024}. A key feature of the NV (and similar systems) is that it features a strong coupling to its host lattice that leads to its quantisation axis being locked along one of the diamond crystal axes. Although this well-defined axis presents opportunities for vector measurements of magnetic fields~\cite{casola2018}, when a strong off-axis field is applied many of the NV's properties degrade~\cite{Tetienne2012}. 
In practice, this has meant that most studies facilitated by NV sensing have either been restricted to materials featuring strong anisotropy\,\cite{tschudinImagingNanomagnetismMagnetic2024, songDirectVisualizationMagnetic2021, Thiel2019, Broadway2020, mclaughlinQuantumImagingMagnetic2022, huangRevealingIntrinsicDomains2023, healey2025} (such that applying a field along the NV axis has a negligible impact on the material) or to weak background fields $\sim1$~mT which do not degrade NV performance regardless of direction\,\cite{liPuzzlingInsensitivityMagnon2023, chenCurrentInducedHidden2024}. 
These restrictions leave a large class of materials and properties that are unable to be comprehensively characterised by NV sensing, including probing transitions that vary under the application of strong fields along different axes. 

An alternative approach is to use the newly discovered optical spin-pair (SP) in hBN (Fig.~\ref{fig:intro}b) and other materials~\cite{scholten2024, robertson2025, gao2024, gao2025, vaidya2025}.
In this case the addressable spin behaves as a spin-1/2 and does not feature a natural quantisation axis with respect to the host lattice. As a consequence the spin response (contrast, resonant frequency) is isotropic with respect to the applied field, circumventing the restrictions placed on NV-like systems (Fig.~\ref{fig:intro}c)\cite{gao2024, scholten2024}. 
Rather than measuring a projection along an existing axis as with the NV, the SP system measures the full scalar field (Fig.~\ref{fig:intro}d).

In this work, we explore how ensembles of SPs can be used in regimes that are not accessible to NV sensors and still obtain useful directional information. 
We show that SPs are robust to extremely large fields and gradients, and we use them to image magnetic transitions in a single crystal of the kagome ferrimagnet TbMn$_6$Sn$_6$~\cite{zhao2025, xu2022, li2023}. 
We image the magnetic fields during the spin-reorientation transition (T$\sim$ 315 K), where TbMn$_6$Sn$_6$ magnetic moments go from lying out-of-plane (OOP) at low temperature to in-plane (IP) at elevated temperature by measuring both the OOP and IP magnetic fields. 
Next, we show how this system can be used to image the IP magnetic fields at any angle. 
Finally, we propose an improved experimental design for probing magneto-crystalline anisotropy in similar materials using the SP system that minimises potential experimental artefacts and shows that meaningful directional information can be inferred despite the sensor's scalar nature.

\begin{figure}
    \centering
    \includegraphics[width=\linewidth]{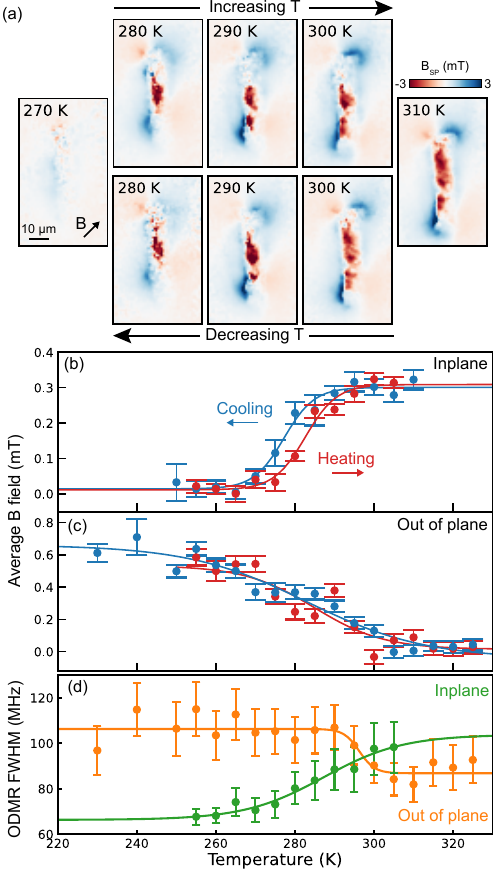}
    \caption{Imaging of a temperature induced spin-reversal transition.
    (a) Exemplary magnetic images of the inplane magnetic field component switching during the spin reversal transition. 
    (b, c) Extracted temperature curves from calculating the average magnetic field around the magnetic material for both (b) IP and (c) OOP fields. Due to the strong magnetic field gradients the regions from directly above the thin-film were not included in the OOP case. 
    (d) The mean full width half maximum of the hBN SP ODMR over the sample as a function of temperature, where the error bars represent the standard deviation. 
    }
    \label{fig:temp}
\end{figure}
\section{Results}
To compare the magnetic imaging capabilities of SPs with NVs we fabricated two devices with a thinned singled crystal of TbMn$_6$Sn$_6$ (150 nm), one with it placed directly on an NV-containing diamond and another on a gold microwave waveguide encapsulated with hBN. 
As the magnetic field from the sample is quite large ($>10$~mT) and oriented away from the NV quantisation axis, the NV spin's contrast is dramatically reduced to the point where no ODMR could be detected near the TbMn$_6$Sn$_6$ film (Fig.~\ref{fig:intro}e)~\cite{Tetienne2012}. 
This can be observed in the large reduction in the ODMR spectra with no signal observed beyond this region.
The remaining signal (orange data, Fig.~\ref{fig:intro}e) is from optical aberrations which gives contribution from NVs far from the pixel imaged~\cite{scholten2022}. The masked region in Fig.~\ref{fig:intro}e (grey) was set by excluding pixels with ODMR contrast $<0.5\%$ where this aberrant feature dominates.
In contrast, the hBN SP is able to image the magnetic field from the material throughout (Fig.~\ref{fig:intro}f).
While it also suffers from the spurious zero field signal it is able to detect the large field above the material ($B\sim10$~mT) which exhibits extreme gradients ($>1$~mT over the 100~nm hBN film) as this local signal does not reduce in contrast.
Imaging the entire field distribution across the sample is important as it helps to capture various inhomogeneities in the material which are not observed in the far field image.
In this case, the inhomogeneity originates from the FIB process used to thin the sample, which damaged the crystal. 

\begin{figure*}
    \centering
    \includegraphics[width=1\linewidth]{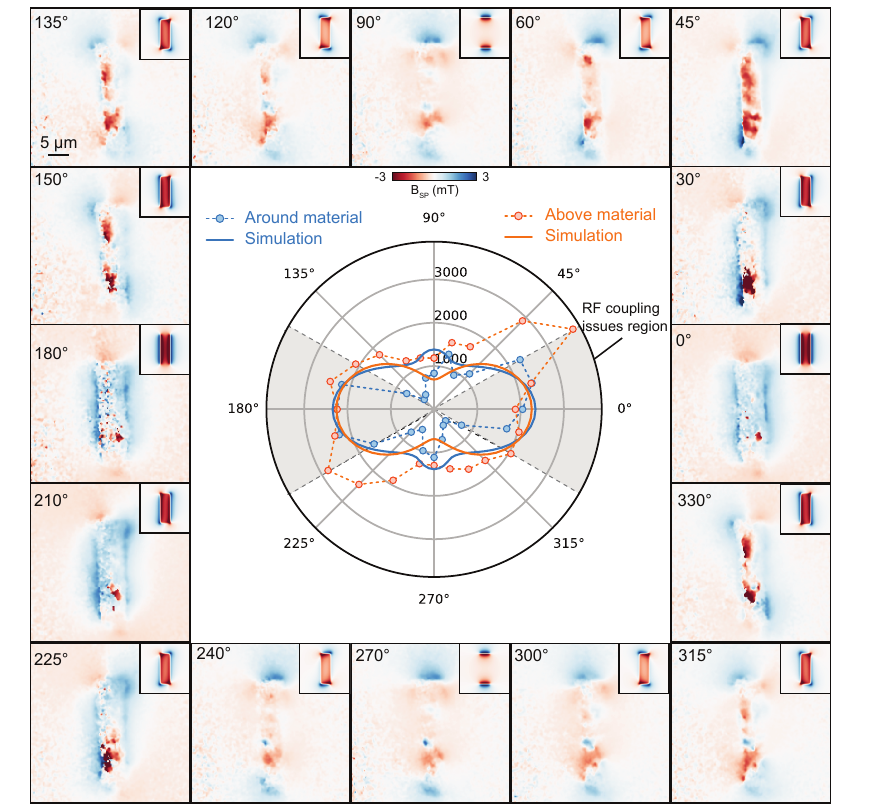}
    \caption{
    Magnetic imaging of in plane anisotropy
    Magnetic images of the inplane magnetic field from the TMS thinned single crystal for various magnetic fields within the plane (taken at $T=315$K). 
    Insert: simulations of the expected magnetic field for a perfect well-ordered thin film with no in-plane crystalline anisotropy. 
    The extracted magnetic field from around the TMS thinned crystal (blue) and above (orange) for the range of magnetic field angles with comparison to the simulation (solid lines). 
    Grey regions around the x-axis correspond to regions were RF to spin coupling is expected to be bad and thus are less reliable. 
    }
    \label{fig:inplane}
\end{figure*}

As a demonstration of the utility of the hBN SP system we first image magnetic fields from a TbMn$_6$Sn$_6$ thinned single crystal through a spin reorientation transition, where the orientation of the magnetisation changes direction from OOP at low temperature to IP in the temperature range of $T= 270 - 330$K~\cite{jones2024, riberolles2022, riberolles2023, riberolles2024, xu2022}.
The origin of this unusual transition is multifaceted, involving energy minimisation of a complex exchange interaction environment that non-trivially evolves as a function of temperature~\cite{riberolles2023, huang2024}.

We use the hBN SP to take a magnetic image at each temperature (Fig.~\ref{fig:temp}a) with an IP applied field of $B = 75$~mT. 
These images can be used to extract the magnetic phase transition by calculating the average magnetic field across the image (Fig.~\ref{fig:temp}(b)).
Interestingly, we observe a slight hysteresis between the heating and cooling curves for an applied in-plane field. This observation could be explained by the applied IP magnetic field helping to stabilise the IP magnetic orientation. 
However, this observation is not significantly above the noise of the experiment and could be driven by probabilistic domain flipping throughout the transition. 
Additionally, we observe that the phase transition occurs significantly below ($T \sim 280$K) the known transition temperature of $T=315$K. 
This discrepancy is due to significant laser heating of the sample which is a known problem in ensemble imaging with solid state spins~\cite{Lillie2020}, but here was deliberately introduced to help overcome the temperature stage limit of $T=325$K in our cryostat.
As laser heating can be difficult to determine directly we make the assumption that the transition occurs at the previously observed temperature and thus suggest that we have laser heating of $T_{\rm heating} \sim 35$K. 

We can also perform the same measurements with an OOP applied field (see Appendix C for images). 
Similar to the IP case, the average magnetic field across the images can be used to access the phase transition. However, in the OOP case there are strong gradients and fields, as discussed previously.
This results in unreliable fitting of the field above the material. 
As such, we introduce a simple mask $B = B(B>0)$, which corresponds to the magnetic fields (blue) that are inherently from regions outside of the material and thus are more reliable.
Interestingly, we observe a very different transition in the OOP case that is significantly broader than the IP case  (Fig.~\ref{fig:temp}b,c) and does not align with a simple two-state phase transition where the probability of being in either state would be conserved.
Instead it points to a complex mixed state where canted magnetisation can locally form as an intermediate phase. 
This is consistent with previous work that has shown Dzyaloshinskii-Moriya interactions can stabilise helical magnetisation during the spin reorientation transition. 
These magnetic helices and the consequent strip domains would be below our spatial resolution~\cite{li2023}. 

Quantum sensors have also been used to explore changes in magnetic noise associated with different spin textures and across phase transitions, similar to susceptibility measurements~\cite{mclaughlin2022, xue2025, ziffer2024}.  
Typically, this involves directly detecting the $T_1$ relaxation or $T_2$ coherence across the phase transition, which when transiting between a ferromagnetic and paramagnetic phase results in a peak in relaxation/coherence rate at the critical temperature. 
As a similar measurement, the SP ODMR width can be used as a proxy to detect changes in low frequency spinwave noise.
Interestingly, the FWHM trends (Fig.~\ref{fig:temp}d) do not directly match those from the stray field signals. Most strikingly, the OOP FWHM shows a sharp change around 295~K while the stray field data changes smoothly over the range of temperatures probed. This indicates that we are not simply measuring field gradients below our spatial resolution and the peak mean FWHM around 100~MHz are related to magentic noise, for instance the slow magnetic fluctuations suggested in previous work~\cite{mielkeiii2022}. The sharp change in the OOP data may in fact be a more precise measure of the spin reorientation temperature, helping to separate reorientation effects from an overall softening of the magnetisation versus temperaure in Fig.~\ref{fig:temp}c.

Importantly, these measurements demonstrate how the flexibility of the imaging platform can to be used to quantify the underlying behaviour more accurately. 
With only a single direction one would incorrectly obtain the critical temperature due to the influence on the applied magnetic field. 
In the case of magnetically driven phase transitions the affect of the magnetic field is implied but during a temperature induced phase transition the effect of the magnetic field changes at each temperature as the effective anisotropy that it is competing against is minimised to zero where the applied magnetic field can now become the dominant energy acting on the magnetisation of the material. 
This positions SP systems as an important tool for more quantitative measurements of temperature-driven phase transitions.  


\begin{figure}
    \centering
    \includegraphics[width=\linewidth]{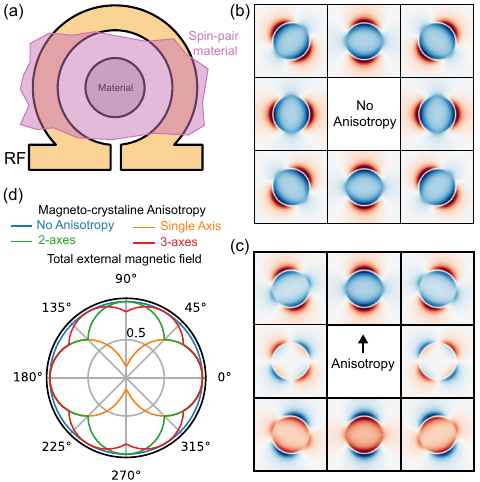}
    \caption{
    Proposal for accurate crystalline anisotropy imaging.
    (a) Illustration of RF delivery for in-plane measurements with a spin-pair material and an optimised material shape for minimising shape based artifacts and anisotropy. 
    (b, c) Simulation of the spin-pair signal with $B=75$~mT and a signal of $B_M=1$~mT for (b) no anisotropy and (c) with a single vertical crystalline anisotropy.  
    (d) Simulated magnetic field from around the material with different degrees of magneto-crystalline anisotropy, assuming instant snapping to the new axis with no dragging. 
    }
    \label{fig:aniso}
\end{figure}

Next we demonstrate how the SP can be used to access an arbitrary magnetic field orientation to explore the magneto-crystalline anisotropy of the TbMn$_6$Sn$_6$ crystal, which is expected to have an easy-plane anisotropy after the spin-reorientation transition and thus should exhibit no preferred orientation within the xy-plane\,\cite{li2023,zhao2025}. 
This measurement utilises the geometric information inherent in stray field imaging to make inferences that would be hidden to e.g. bulk transport measurements.
To access this we take magnetic images of the TbMn$_6$Sn$_6$ thinned single crystal while varying the applied magnetic field angle within the plane at every 15$^\circ$ with a strength of $B=75$~mT, a subset of which is shown in Fig.~\ref{fig:inplane}.
Just like in the temperature example we can use the average magnetic field in the image to extract the behaviour of the material to these magnetic field changes (Fig.~\ref{fig:inplane} inner panel). 
In this case there is a different behaviour above the material to the outside of the material, which can be separated through comparing different regions of interest or masking. 
We perform numerical simulation assuming zero anisotropy of the magnetic field from a similar rectangle (Fig.~\ref{fig:inplane} inserts) and calculate the same averages. 
This shows how the shape of the material modifies the average field itself and deviations from these shape would need to be accessed to identify anisotropy.
Our experimental results are qualitatively in agreement with the simulations -- any deviation is too subtle to recognise given the level of noise in the measurements. 

The interpretation of our measurements is limited to qualitative inferences in this case for two reasons that we have already identified. Firstly, the stray field patterning is a result of the material's shape as well as its magnetisation direction (even before the possibility of shape anisotropy is considered). 
In the NV literature, ambiguities in angular response have been addressed using machine learning techniques but these take advantage of the projective nature of measurement along the well-defined NV quantisation axis~\cite{Broadway2025, tschudin2024, healey2025}. 
It is unclear whether a scalar sensor with a reorientable axis (Fig.~\ref{fig:intro}d) is capable of producing unique enough images to facilitate such analysis. 
Secondly, measurements made in weak microwave (MW) driving geometries are less reliable, preventing the realisation of a truly isotropic magnetic sensor. 

We explore whether these limitations can be circumvented by considering an ideal measurement scenario, sketched in Fig.~\ref{fig:aniso}a. Here the MW delivery is perfectly oriented along the out-of-plane axis (for instance facilitated by an Omega-shaped waveguide), such that the driving of the SP will be equivalent for applied fields along any in-plane direction. 
To remove the impact of the magnetic material's geometry on the measurement we make it circular. 
Fig.~\ref{fig:aniso}b and c show simulations of the stray field signal we would expect to measure with and without a strong uniaxial magnetocrystalline anisotropy respectively, rotating the measurement field within the plane. 
In this extreme case there are clear qualitative differences between the two cases. 

As in Fig.~\ref{fig:inplane} we can extract the net fields extending beyond the magnet as a simple figure of merit. 
Simulated profiles are shown in Fig.~\ref{fig:aniso}d, where we include the 2-axis and 3-axis magnetocrystalline anisotropy cases as well as the uniaxial and zero anisotropy cases. We can see that as the deviation from the zero-anisotropy case becomes more subtle as we move to higher order symmetries, but that there is still a clear lobe structure even in the 3-axis case that could potentially be resolvable in experiment. Although the experimental realisation is beyond the scope of the current work, these simulations demonstrate the possibility for extracting meaningful directional information using the SP system. 

\section{Conclusion}
We have shown that the SP system in hBN offers distinct imaging capabilities to traditional solid state quantum sensors such as the NV in diamond by virtue of its fully isotropic response to magnetic field. 
Specifically, we were able to explore the properties of a strongly magnetic sample that compromised the operation of an NV sensor. 
As a proof of principle, we tracked the evolution of a spin reorientation transition along two orthogonal axes as well as the rotation of magnetisation throughout the XY plane. 
In practice, the true extent of directional information that can be obtained in a given experiment may be limited by factors such as the orientation of the MW drive and sample shape, but we have shown that these can be mitigated. 
These results establish SP systems in hBN and other materials as versatile magnetic imaging platforms.
\section*{Acknowledgements}
This work was supported by the Australian Research Council through grants CE200100010, DE230100192, DP220103783, DP250100973. This work was performed in part at the Melbourne Centre for Nanofabrication (MCN) in the Victorian Node of the Australian National Fabrication Facility (ANFF). The FIB fabrications are conducted in Monash Centre for Electron Microscope (MCEM), Monash University, a Microscopy Australia (ROR: 042mm0k03) facility supported by NCRIS. The FIB equipment is funded by Australian Research Council grants (LE200100132). HHT acknowledges the ANFF ACT Node for providing access to the facility for the growth of hBN.

\appendix
\section{Sample Fabrication}
\begin{figure}[h]
    \centering
    \includegraphics{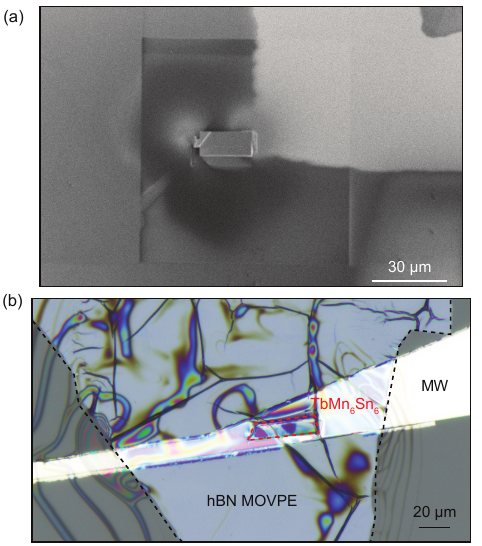}
    \caption{(a) Scanning electron microscope image of TbMn$_6$Sn$_6$ film that was placed on the diamond in Fig.~\ref{fig:intro} after the focused ion beam removal the surrounding material. 
    (b) Optical image of the TbMn$_6$Sn$_6$ film encapsulated with the hBN MOVPE sensing layer used throughout the manuscript.}
    \label{sifig: sample fabrication}
\end{figure}

The high quality TbMn$_6$Sn$_6$ single crystals grown from Sn-flux method following the same procedure in ref.\,\cite{yin2020} were employed as starting materials. Using a focused ion beam (FIB) technique equipped in
a dual beam scanning electron microscope (SEM, Thermofisher FEI Helios 5), single crystals of 150 nm were prepared and transferred to the hBN or diamond devices. Fig.~\ref{sifig: sample fabrication}a shows the isolation of a TbMn$_6$Sn$_6$ section from the bulk single crystal using the FIB and Fig.~\ref{sifig: sample fabrication}b shows such a section transferred onto a MW stripline and encapsulated with hBN for measurement.

\section{Experimental Apparatus}
The experiments were performed inside a closed-cycle cryostat with a base temperature of 4K (AttoDry1000) with a 1-T superconducting vector magnet (Cryomagnetics).
The optical excitation of the hBN and dimaond material was performed with a 532~nm continuous wave laser (Laser Quantum Ventus 1~W, coupled to a single-mode fibre and a 60~mm collimation lens at the output of the fibre was used to adjust the beam size at the sample. 
The laser was directed onto the sample through a dichroic beam splitter and a low-temperature microscope objective (Attocube LT-APO/VISIR/0.82).
The photolumenscence from the quantum sensors was collected using a 4f lens configuration and passes though the dichroic mirror and a 731/137~nm band pass filter, before being focused (300~mm tube lens) onto a water cooled sCMOS camera (Andor Sona).
The microwave control was performed with a signal generator (Rohde \& Schwarz SMB100A), a switch (Mini-Circuits ZASWA-2-50DR+), and a 50~W amplifier (Mini-Circuits HPA-50W-63), which is directed into the cryostat and through a custom made printed circuit board containing a straight waveguide with a width of 1~mm.

\section{Additional data}
\begin{figure}[h]
    \centering
    \includegraphics[width=1\linewidth]{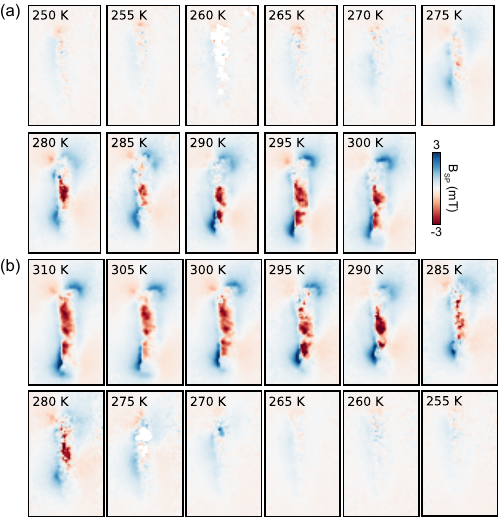}
    \caption{Magnetic images from hBN SPs of the TbMn$_6$Sn$_6$ spin reversal with an IP magnetic field corresponding to the data in Fig.~\ref{fig:temp} for both (a) increasing and (b) decreasing temperature.}
    \label{sifig: temp IP}
\end{figure}

\begin{figure}[h]
    \centering
    \includegraphics[width=1\linewidth]{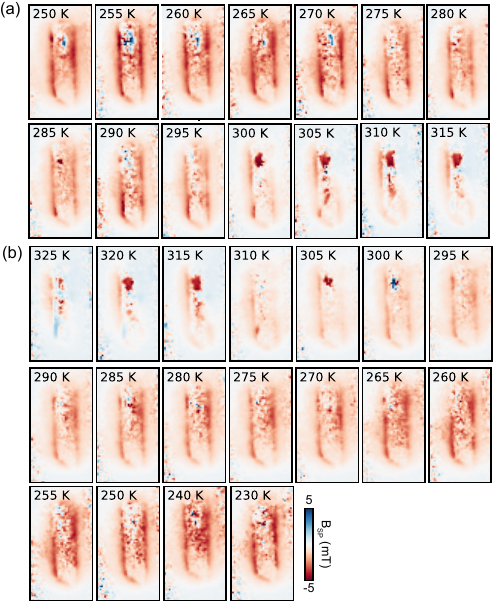}
    \caption{Magnetic images from hBN SPs of the TbMn$_6$Sn$_6$ spin reversal with an OOP magnetic field corresponding to the data in Fig.~\ref{fig:temp} for both (a) increasing and (b) decreasing temperature.}
    \label{sifig: temp OOP}
\end{figure}

We have included the datasets of the extracted data in Fig. 2 in Fig.~\ref{sifig: temp IP} and Fig.~\ref{sifig: temp OOP}.

\bibliographystyle{naturemag}
\bibliography{bib.bib}

\end{document}